\documentclass{aastex62}

\begin{document}

\title{Mapping Milky Way Halo Substructure using Stars in the Extended Blue Tail of the Horizontal Branch}

\author{Robert Gryncewicz}
\affiliation{Department of Physics, Applied Physics and Astronomy, Rensselaer Polytechnic Institute, Troy, NY 12180, USA}

\author{Heidi Jo Newberg}
\affiliation{Department of Physics, Applied Physics and Astronomy, Rensselaer Polytechnic Institute, Troy, NY 12180, USA}

\author{Charles Martin}
\affiliation{Department of Physics, Applied Physics and Astronomy, Rensselaer Polytechnic Institute, Troy, NY 12180, USA}

\author{Thomas Donlon II}
\affiliation{Department of Physics, Applied Physics and Astronomy, Rensselaer Polytechnic Institute, Troy, NY 12180, USA}

\author{Paul Amy}
\affiliation{Department of Physics, Applied Physics and Astronomy, Rensselaer Polytechnic Institute, Troy, NY 12180, USA}

\correspondingauthor{Heidi Jo Newberg}
\email{heidi@rpi.edu}

\begin{abstract}

Although Blue Horizontal Branch (BHB) stars are commonly used to trace halo substructure, the stars bluer than $(g-r)<-0.3$ are ignored due to the difficulty in determining their absolute magnitudes. The blue extention of the horizontal branch (HBX) includes BHB tail stars and Extreme Horizontal Branch (EHB) stars.
We present a method for identifying HBX stars in the field, using spectra and photometry from the Sloan Digital Sky Survey Data Release 14 (SDSS DR14). 
We derive an estimate for the absolute magnitudes of BHB tail and EHB stars as a function of color, and use this relationship to calculate distances.
We identify an overdensity of HBX stars that appears to trace the northern end of the Hercules-Aquila Cloud (HAC).
We identify three stars that are likely part of a tidal stream, but this is not enough stars to explain the observed overdensity.
Combining SDSS data with \textit{Gaia} DR2 proper motions allows us to show that the orbits of the majority of the HBX stars in the overdensity are on high eccentricity orbits similar to those in the Virgo Radial Merger/\textit{Gaia}-Enceladus/\textit{Gaia} Sausage structure, and that the overdensity of high eccentricity orbits extends all the way to the Virgo Overdensity. We use stellar kinematics to separate the HBX stars into disk stars and halo stars. The halo stars are primarily on highly eccentric (radial) orbits. The fraction of HBX stars that are EHBs is highest in the disk population and lowest in the low eccentricity halo stars.

\end{abstract}

	
\section{Introduction}
When dwarf galaxies fall into the Milky Way galaxy, tidal forces pull them apart and their stars become bound to the Milky Way rather than the parent dwarf galaxy. If the dwarf galaxy falls in on a circular orbit, a tidal stream of stars is created, with each star following an orbit similar to the dwarf galaxy progenitor's orbit. Stars that are stripped with lower energy per mass have shorter orbital periods and are pulled ahead, and stars with higher energy per mass lag behind. The entire dwarf galaxy can be dissolved into a long ribbon of stars that encircles the Galaxy. See \citet{Newberg_Carlin2016}, and chapters therein, for a review of the disruption process and properties of the first couple dozen tidal streams that were discovered.

Dwarf galaxies that fall into a more massive galaxy on radial orbits also disrupt. However, instead of forming tidal streams, they catastrophically disrupt on the first pass through the galaxy center. The stars orbit the host galaxy on radial orbits, piling up at apogalacticon on umbrella-shaped ``shells" before falling back through the galaxy center \citep{Hernquist1988, sanderson2013}. Stars from the dwarf galaxy will follow orbits with a range of energies, and therefore a range of periods; the Galactocentric distance and number of the umbrella-shaped shells will vary with time as stars of different energy reach apogalacticon.

The recent release of {\it Gaia} DR2 \citep{Gaia_Mission,Gaia_DR2} led to the discovery of the \textit{Gaia} Sausage merger \citep{bel_sausage}, the {\it Gaia}-Enceladus merger \citep{Helmi_enc}, and the Virgo Radial Merger \citep[VRM;][]{tom_paper}, which were discovered in different Milky Way locations by different techniques. Although the inferred merger timescale differs between different authors, the similarity of the orbits of the stars in each claimed merger event suggests that they originate from the same dwarf galaxy merger event. The relationship might not be quite one-to-one, however; there is a possibility that the {\it Gaia}-Enceladus merger as discovered includes debris from both the {\it Gaia} Sausage and at least one other halo substructure combined \citep{biggest_splash}. The {\it Gaia} Sausage \citep{2019MNRAS.482..921S} and VRM \citep{tom_paper} have both been shown to produce a significant amount of stellar debris in the region of the Virgo Overdensity \citep[VOD;][]{vivas, 2008ApJ...673..864J} and the Hercules-Aquila Cloud \citep[HAC;][]{belokurov2007}, where the stars reach apogalacticon. This radial merger event is thought to have contributed the majority of the stars in the Milky Way inner halo \citep{2013ApJ...763..113D, bel_sausage, 2019MNRAS.490.3426D}, and has changed the way we think about searching for substructure in the outer parts of our galaxy. Recently, \citet{tom_new_paper} identified shell structure in the VOD and HAC, where the stellar halo has been known to be overdense, using blue horizontal branch (BHB) and RR Lyrae stars as tracers. 

This paper explores the use of the extended blue tail of the horizontal branch, including Extreme Horizontal Branch stars (EHBs), to map structure in the Milky Way galaxy. \cite{1974ApJS...28..157G} used the terminology Extended Horizontal Branch (EHB) stars, which conflicts with the current common practice of using EHB for Extreme Horizontal branch stars, which inhabit the extreme blue end of the blue tail. We will refer to stars on the blue extension of the horizontal branch as Horizontal Branch Extension (HBX) stars. 

Previous maps of the Milky Way's stellar halo limited the color range of BHBs to $-0.3<g-r<0.0$ \citep[e.g., ][]{2000ApJ...540..825Y,2004AJ....127..899S}. This is because horizontal branch stars that are bluer than $g-r=-0.3$ become fainter in the $g$ filter and form a ``shoulder" where the horizontal branch turns downwards towards bluer stars, as the absolute magnitude, $M_g$, increases (gets fainter). \citet{2011MNRAS.416.2903D} derived a polynomial fit to get a more accurate fit to the absolute magnitude of horizontal branch stars as a function of color, but still limited the fit to $-0.25<g-r<0.0$.

The connection between the bluest horizontal branch stars (which are most easily identified as bluer than the shoulder of the horizontal branch in globular clusters) and blue subdwarfs in the field has been studied for many years.  \citet{1974ApJS...28..157G} suggested that the blue end of horizontal branch stars extended from spectral type A to B and all the way to O, and explained the origin of the blue (sdB and sdO) halo that they observed. \citet{1986A&A...162..171H} showed that spectroscopically the field stars and the the corresponding EHB stars in globular clusters were identical. However, a solid understanding of the mechanism through which the EHB stars are produced has remained elusive. \citet{2009ARA&A..47..211H} gives a review of the field of hot subdwarfs and the struggle to match them to their evolutionary origins.

More recently, the morphology of the horizontal branch has begun to be better understood. \citet{1976ApJ...204..804N} showed that there were two gaps (and therefore three distinct populations) of high latitude faint blue stars, that were expected to be part of this extended horizontal branch; the gaps were measured from Str{\"o}mgren {\it ubvy} photometry to be at $\log T_{\rm eff} \sim 4.11$ and $\log T_{\rm eff}  \sim 4.34$. These gaps are similar to the Grundahl jump \citet{1998ApJ...500L.179G} discovered in M13 in the Str{\"o}mgren {\it uvby}$\beta$ system, and the Momany jump \citep{2002ApJ...576L..65M} discovered in NGC 6752 in $U$ vs. ($U-V$). These gaps are described below.

\citet{1998ApJ...500L.179G} showed that bluer than $(u-y)_0=0.95$, the horizontal branch is 0.4 magnitudes brighter in $u$ than zero age horizontal branch models would predict. The authors suggested that this would indicate two populations on the horizontal branch. The anomalous bluer horizontal branch stars could be caused by He mixing in the pre-horizontal branch giants, or variations in the CNO abundances. In a subsequent paper, \citet{1999ApJ...524..242G} show that this jump is ubiquitous in globular clusters that have a horizontal branch that extends hotter than $T_{\rm eff}\sim 11,500$ K. Because the jump does not depend on properties of the cluster (such as metallicity) and is also seen in the field (as a jump in surface gravity), the later paper suggests that there is no evolutionary difference between the stars redder and bluer than the Grundahl jump, but instead the stellar atmosphere models need to be adjusted to fit the data. \citet{1999ApJ...524..242G}, \citet{2006A&A...452..493P}, and \citet{2016ApJ...822...44B}, and references therein, explore radiative levitation of metals and gravitational settling of helium to explain the Grundahl jump.

\citet{2002ApJ...576L..65M} showed that in addition to the Grundahl jump, there is a jump at $U-V \sim -1.0$, that was attributed to post-horizontal branch evolution and/or diffusion effects. \citet{2004A&A...420..605M} showed that this second jump (and possibly a third) was common in clusters with horizontal branches that extended blue enough to populate both sides of the jump. The third gap lies between the EHB stars and the “blue-hook” stars, also known as “hot He-flashers” \citep{1996ApJ...466..359D}. \citet{2016ApJ...822...44B} analyze photometry for 53 globular clusters, and find that these jumps are remarkably consistent, at the same temperature and color, between the clusters. Because these clusters have a range of stellar populations, the consistency of the jumps suggests that they are a result of changes in the stellar atmospheres of core helium fusing stars at different temperatures. Although extreme differences in He abundance were found to shift the Grundahl jump in two cases, the Momany jump remained stable for all clusters with a sufficiently hot blue horizontal branch. \citet{2017ApJ...851..118B} showed that the stars hotter than the Momany jump had lower Fe in their atmospheres, which could also be related to atmospheric affects such as radiative acceleration.

We will show that the Grundahl jump is close to $(g-r)=-0.3$, the bluest color for which horizontal branch stars have been used to map the Milky Way halo. The Momany jump, located at $\sim 20,000$ K, can be identified within the sample of stars we select, and nominally separates stars that we will call ``BHB tail" from EHB stars. Since there are very few blue-hook stars compared to other stars on the HBX, their presence or absence in our dataset makes little difference to our result. Note that for our purposes it is not important to understand why stars land at one temperature or another on the horizontal branch or HBX; it is only important that these stars are present in particular populations of stars, and that the distance can be roughly estimated from the colors. We show that distances to HBX stars can be found to about 20\%, which is considerably better than distance measurements to turnoff stars, which have also been used to trace density substructure in the halo \citep{2002ApJ...569..245N}.

We were inspired to search for EHB stars as halo tracers because it was suggested by \citet{scibelli_paper} that the very blue stars which match the Sloan Digitial Sky Survey (SDSS) DR14 A0p template might be EHB stars. After exploring the relationship between SDSS spectral type and color, we determined that most of the A0p stars identified by the SDSS pipeline were probably BHB tail stars. However, we recovered a larger number of EHB and BHB tail stars by including all of the blue stars that were not white dwarfs, main sequence stars, or subdwarf stars in the sample. The analysis we present does not restrict the stellar types of the horizontal branch stars determined from template matching, but does use the template matching to remove stars that are not horizontal branch stars.

We show how field EHB and BHB tail stars can be identified, provide a function with which to estimate the absolute magnitude (and therefore the distance) to each star, and then use these stars to explore the substructure of the Milky Way halo. We find a large number of HBX stars in the HAC that have similar orbits to those of stars in the VRM, and we observe an overdensity of halo stars at low Galactic latitude that stretches from the HAC to the VOD.

\section{Selecting HBX Stars}
\subsection{Sample Selection}
We first construct a sample of very blue point sources that are not white dwarf stars, main sequence O or subdwarf O stars, as determined by their spectrum.
Spectroscopically observed blue stars with an extinction-corrected color of $(g-r)_0 < -0.3$ were selected from 
the Sloan Digital Sky Survey (SDSS) Data Release 14 \citep[DR14,][]{sdss}.
This color cut is slightly more restrictive than the $(g-r)_0 < -0.25$ cut used by \citet{scibelli_paper} in their analysis.
Data was also restricted to only those stars that could be matched with \textit{Gaia} DR2 \citep{gaia_cite}, so we have full 6D information for all of the stars in our sample.
We then removed stars with spectral types related to white dwarfs, as determined from SDSS DR14 template matching. 
A more complete exploration of the SDSS templates for blue stars is given by \citet{scibelli_paper}, who 
determined that the subclass OB and O objects were likely dwarf and sdO stars;  these stars were removed from our dataset as well.
The other sample stars consist primarily of spectral types A0p (389), A0 (225), B9 (246), A9V (231) and B6 (135). The remaining stars (353) have other spectral types primarily related to O/B/A stars, but some have redder spectral types that indicate they were misclassified. In total we select 1579 stars across the SDSS footprint. 

\subsection{Overdensity}
\begin{figure}
\centering
\includegraphics[width = \linewidth]{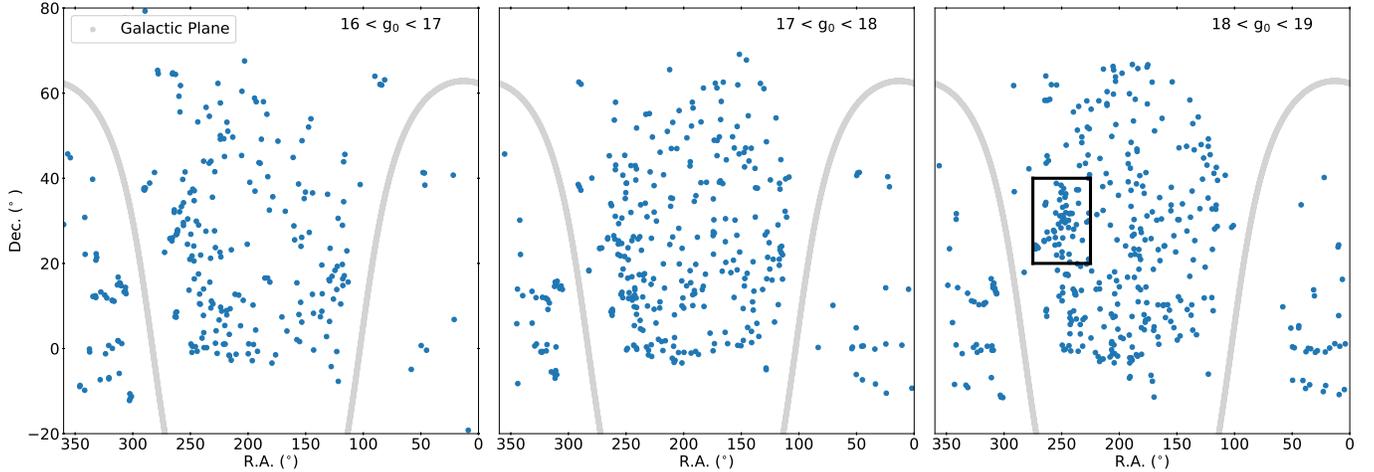}
\caption{
Sky position for the candidate HBX stars at different apparent magnitude ranges. Stars were selected with reddening corrected color of $(g - r)_0 < -0.3$. 
In the right-most figure there is an apparent overdensity spanning $225^\circ < \rm{R.A.} < 275^\circ$ and $20^\circ < \rm{Dec.} < 40^\circ$. This region of the sky does not have any apparent overdensity in the 17 to 18 magnitude range (middle). In the 16 to 17 magnitude range (left) there are stars present in this area of the sky, but it is not higher density than adjacent sky areas with similar Galactic latitude. The gray lines show the position of the Galactic plane.
}
\label{fig:ra_dec_plots}
\end{figure}

Figure \ref{fig:ra_dec_plots} shows the sky positions of our presumed HBX stars for different apparent magnitude ranges.
Two thirds of the total number of stars in the data selection are included in the $16 < g_0 < 19$ magnitude range shown in the figure.
The footprint of the SDSS survey is evident in the diagram. The majority of the stars with spectra are in the north Galactic cap. Especially for the brighter magnitudes, there is a clear increase in the star counts towards the Galactic plane.
Note the apparent overdensity that is highlighted by the black box in the $18 < g_0 < 19$ panel of Figure \ref{fig:ra_dec_plots}.
This overdensity is located in the same direction as the northern end of the HAC region (R.A., Dec.) = ($250^\circ$, $30^\circ$), and is not present in other magnitude ranges. 

\begin{figure}
\centering
\includegraphics[width =\linewidth]{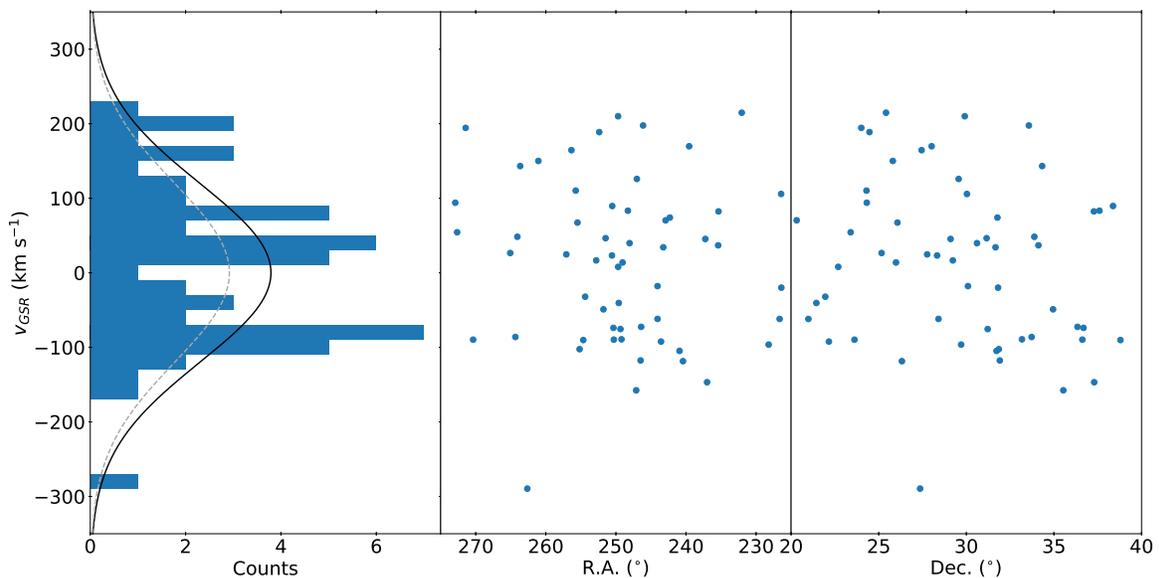}
\caption{
Line-of-sight velocities of stars in the 18 $<$ g$_0$ $<$ 19 overdensity shown in Figure \ref{fig:ra_dec_plots}. The left panel shows a histogram (bin size of 20 km s$^{-1}$) of $V_{GSR}$ compared to a Gaussian centered on 0 km s$^{-1}$ with a dispersion of 120 km s$^{-1}$, normalized to the star count. 
We see an excess at $V_{GSR}= -90$ km s$^{-1}$, and a less significant second excess at $V_{GSR}=+30$ km s$^{-1}$. 
If we remove these two apparent peaks before normalizing the background we obtain the gray, dashed Gaussian fit. In the middle and right panels the velocities are plotted as a function of R.A. and Dec., respectively. 
The -90 km s$^{-1}$ peak has an excess of 7-8 stars over the 4-5 background counts (12 total). If the background is normalized with the excess counts from the two peaks removed (gray dashed line), then the -90 km s$^{-1}$ peak has an excess of 8 stars over the 4 expected background counts.
}
\label{fig:vgsr_ra}
\end{figure}

Figure \ref{fig:vgsr_ra} shows the line-of-sight, Galactic standard of rest velocity, $V_{GSR}$, of the stars in the black box of Figure \ref{fig:ra_dec_plots}.
In this figure, the distribution of $V_{GSR}$ for the stars is compared to a background Gaussian centered at 0 km s$^{-1}$ with a dispersion of 120 km s$^{-1}$, which is characteristic of generic halo stars and is normalized so that the area under the Gaussian integrates to the 57 stars in the selection box. However, the velocity distribution of stars does not look very Gaussian. It looks like there are at least one or two peaks above any Gaussian background: one is located at $-90$ km s$^{-1}$ and the other is at $+30$ km s$^{-1}$. If the $V_{GSR}$ peaks are substructure, then they do not belong to the background halo and we should not include them in the normalization of the underlying Gaussian distribution. In order for the Gaussian background plus the excess counts above the background in each of these peaks to sum to the 57 stars observed, the background Gaussian amplitude must be decreased by a factor of 0.77 (dashed line in the left panel of Figure~\ref{fig:vgsr_ra}). For this case, there are 12 stars where we expect 4.3 stars in the $V_{GSR}=-90$ km s$^{-1}$ peak (3.7 sigma), and 11 stars where we expect 5.5 stars in the $V_{GSR}=+30$ km s$^{-1}$ peak (2.3 sigma). This calculation is meant to give a rough number of stars expected in each ``overdensity," and the exact statistics are not important. For example, in the more significant peak, we could hope to find a substructure with 8 stars. 
The HBX stars in these overdensities are in the color range $-0.53 < (g-r)_0 < -0.34$.
The middle and right panels of Figure \ref{fig:vgsr_ra} show the relationship between $V_{GSR}$ and sky position.
While the $+30$ km s$^{-1}$ stars are found spread across the selected range of right ascensions, the stars near $V_{GSR} = -90$ km s$^{-1}$ are possibly concentrated near (R.A., Dec.) $= (248^\circ, 33^\circ)$. However, a tight group of 8 stars is not apparent.

\section{HAC Velocities}
We first ask whether either of the $V_{GSR} = -90$ km s$^{-1}$ or $+30$ km s$^{-1}$ velocity peaks are consistent with the previously published velocity of the HAC.
\citet{belokurov2007}, hereafter B07, measured a $V_{GSR}$ of $+180$ km s$^{-1}$ for the northern end of the HAC, using data from a preliminary version of SDSS DR5.
This velocity is not reproduced with newer data releases \citep{martin2016}, as we will now show.
We selected stars in the HAC region, $20^\circ < l < 75^\circ$ and $20^\circ < b < 55^\circ$, with $r_0 < 19$ and $0.0 < (g-r)_0 < 1.0$. 

\begin{figure}
\includegraphics[width=\linewidth]{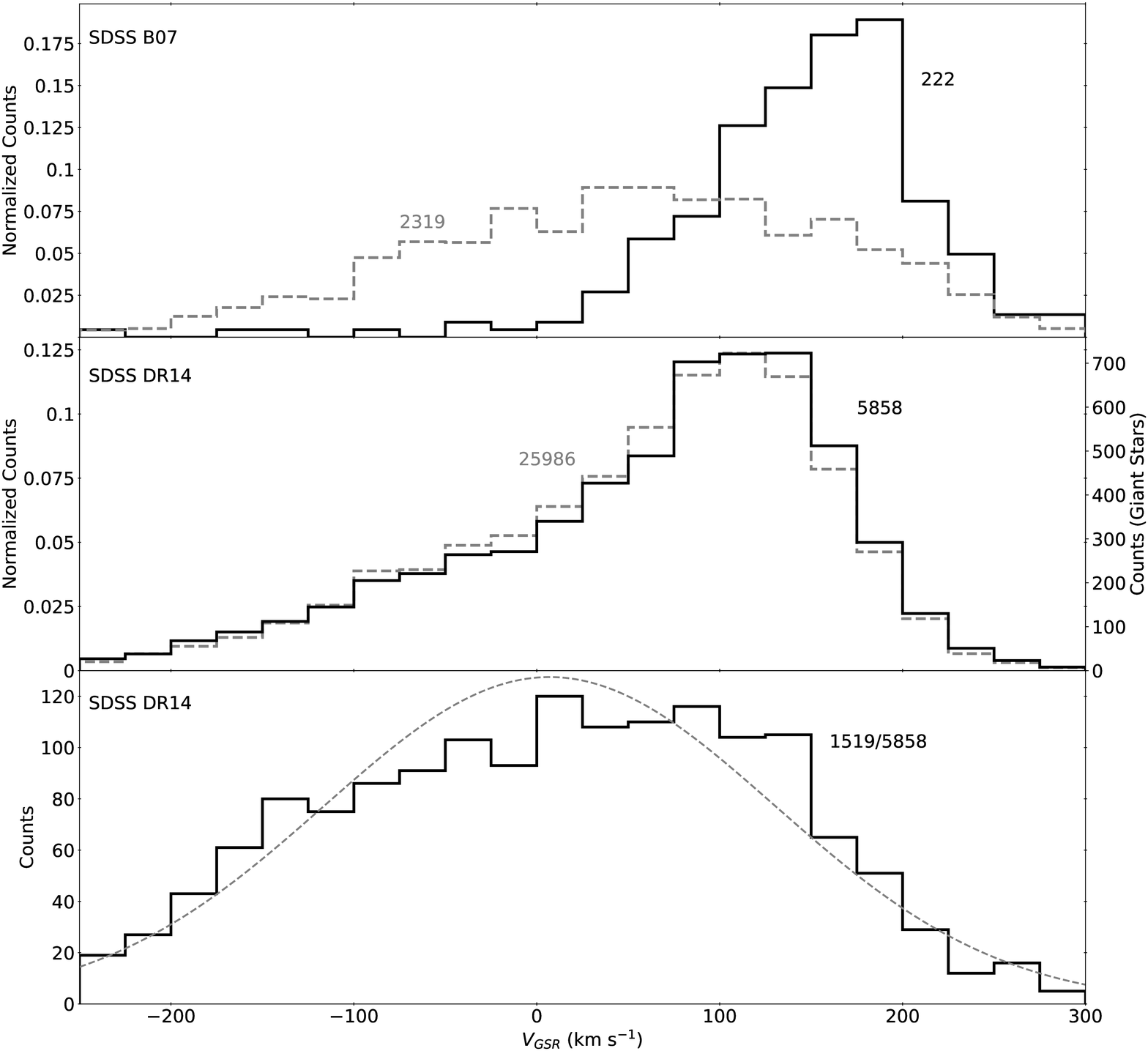}
\centering
\caption{
The velocity in the HAC region with SDSS DR14 data, recreated from \citet{martin2016}.
The top panel shows the original data from \citet{belokurov2007}, from which the velocity of the northern part of the HAC was determined to be $+180$ km s$^{-1}$. The field data was selected from a pre-DR5 version of the SDSS data in the region $20^\circ < l < 75^\circ$ and $20^\circ < b < 55^\circ$, with cuts for $r_0 < 19$ and $0.0 < (g-r)_0 < 1.0$. 222 candidate RGB stars (solid line) were selected from the field between M92's ridgelines, shifted to 10 and 20 kpc. The field, non-RGB star sample (dashed) consisted of 2319 stars; although there are a few more positive velocity stars than negative velocity stars, the velocity distribution contains no clear peaks. The difference between presumed main sequence and giant stars led the B07 authors to conclude that the RGB selection was successful and the HAC $V_{GSR}$ was +180 km s$^{-1}$.
The middle panel shows the same cuts but instead applied to stars in SDSS DR14. The DR14 data has a peak of $V_{GSR} = +120$ km s$^{-1}$ in both the selected giants and the field stars. This peak is consistent with the expectations for main sequence stars in thick disk in this direction, supporting the idea that the original photometric RGB selection was not clean. Further evidence is found in the bottom panel. A surface gravity cut was applied to the DR14 data to remove thick disk contamination, reducing the sample from 5858 to 1519 stars. The bottom panel shows that a (mostly) pure sample of halo giant stars has a Gaussian distribution centered around 0 km s$^{-1}$ as expected for a generic halo population. This suggests that the published HAC velocity is a product of incorrect velocity values in the preliminary version of DR5 data.
}
\label{fig:vgsr_hist}
\end{figure}

Then, following the method outlined in B07, candidate giant stars were identified by selecting only stars between two 
M92 ridgelines \citep{clem2005}, shifted to 10 and 20 kpc.
Though the majority of the field stars are expected to be main sequence stars, the selection between the ridgelines favors stars at the distance of the HAC.
The top panel of Figure \ref{fig:vgsr_hist} shows the original results from B07, using the same data that was used for that paper.  The fact that 222 candidate HAC RGB stars have a different velocity distribution gives the impression that the selection of the HAC RGB stars was successful, and that the velocity peak indicates the line-of-sight velocity of the HAC. However, we show below that the peak was just an artifact of early data pipelines.

The middle panel of Figure \ref{fig:vgsr_hist} shows the result of the
same selection, but with a newer data release: SDSS DR14. In DR14, there are many more stars present than were originally available in DR5. The newer data selection includes 5858 photometrically selected candidate giant stars, out of roughly 26,000 stars in the field. In the newer data, the $V_{GSR}$ histograms for field and giant stars no longer differ from each other. The velocity of the peak at $120$ km s$^{-1}$ is consistent with the assumption that most of the stars in both the field and candidate RGB stars are main sequence thick disk stars. A comparison of the velocities of individual stars in the B07 dataset to the velocities of the same stars in SDSS DR14 shows that some of the values in the original dataset were erroneous and resulted in the shifted location of the peak in the top panel of Figure \ref{fig:vgsr_hist}. These erroneous velocities were corrected in later data analysis pipelines.

We further corroborate this point by utilizing SDSS DR14 surface gravity information, which was also not available in DR5. 
Imposing a surface gravity cut of $0.0 < $ log g$_{\rm{ADOP}}$ $ < 3.5$  removed most of the main sequence thick disk stars from the sample; the sample was reduced from 5858 stars to 1519 stars. The bottom panel of Figure \ref{fig:vgsr_hist} shows the line-of-sight velocity histogram for the cleaner dataset. The relatively smooth Gaussian distribution is consistent with our expectation for halo stars. Evidently, even the nearly pure sample of giant stars does not give us an indication of a group motion of the stars in the HAC.

In \citet{tom_paper} and \citet{tom_new_paper}, the VRM was shown to be connected to the HAC. \citet{tom_new_paper} showed that this merger generated caustic structures with stars moving back and forth through the center of the Milky Way along radial orbits. 
Stars with lower energy are more tightly bound to the Milky Way and oscillate with higher frequency. When the stars reach apogalacticon, they have a radial velocity of 0 km s$^{-1}$ with respect to the Galactic center. Closer to the Galactic center but in the same caustic structure, some of the stars will be moving towards the Galactic center and some will be moving away. The range of different velocities of stars in the same part of the Galaxy produces a complex line-of-sight velocity distribution as seen from the Solar position, especially when stretched over several distance ranges. 
Therefore the stars in the HAC would not be expected to have similar velocities; the velocity of each star in the overdensity is dependent on its own individual radial orbit. 
Viewing the velocity structure in Figure~\ref{fig:vgsr_ra} through this lens, it seems plausible that we are seeing both positive and negative line-of-sight velocity peaks from a radial merger event. Similar velocity structures have been identified in the VOD, and are indicative of radial mergers.

\section{Distances to HBX Stars}
\subsection{Identifying Known HBX Stars}

In order to compute 6D information, we need to calibrate distances to the stars in our sample using the properties of known HBX stars in globular clusters. We used photometric data from 
\citet{daophot_ref}; their DAOPHOT reductions of SDSS imaging data for globular clusters allowed for more accurate crowded field photometry than was obtained with the regular SDSS pipeline.

Spectra in the vicinity of six clusters that had a blue horizontal branch that turned fainter on the blue end and were within the footprint of SDSS were collected from 
\citet{2011_smolinski} and \citet{2016AJ....151....7M}. 
They used the Sloan Extension for Galactic Understanding and Exploration \citep[SEGUE;][]{2009AJ....137.4377Y} data, and determined cluster membership using radial velocities and several other known properties of the globular clusters. 

\begin{figure}
\centering
\includegraphics[width = .98\linewidth]{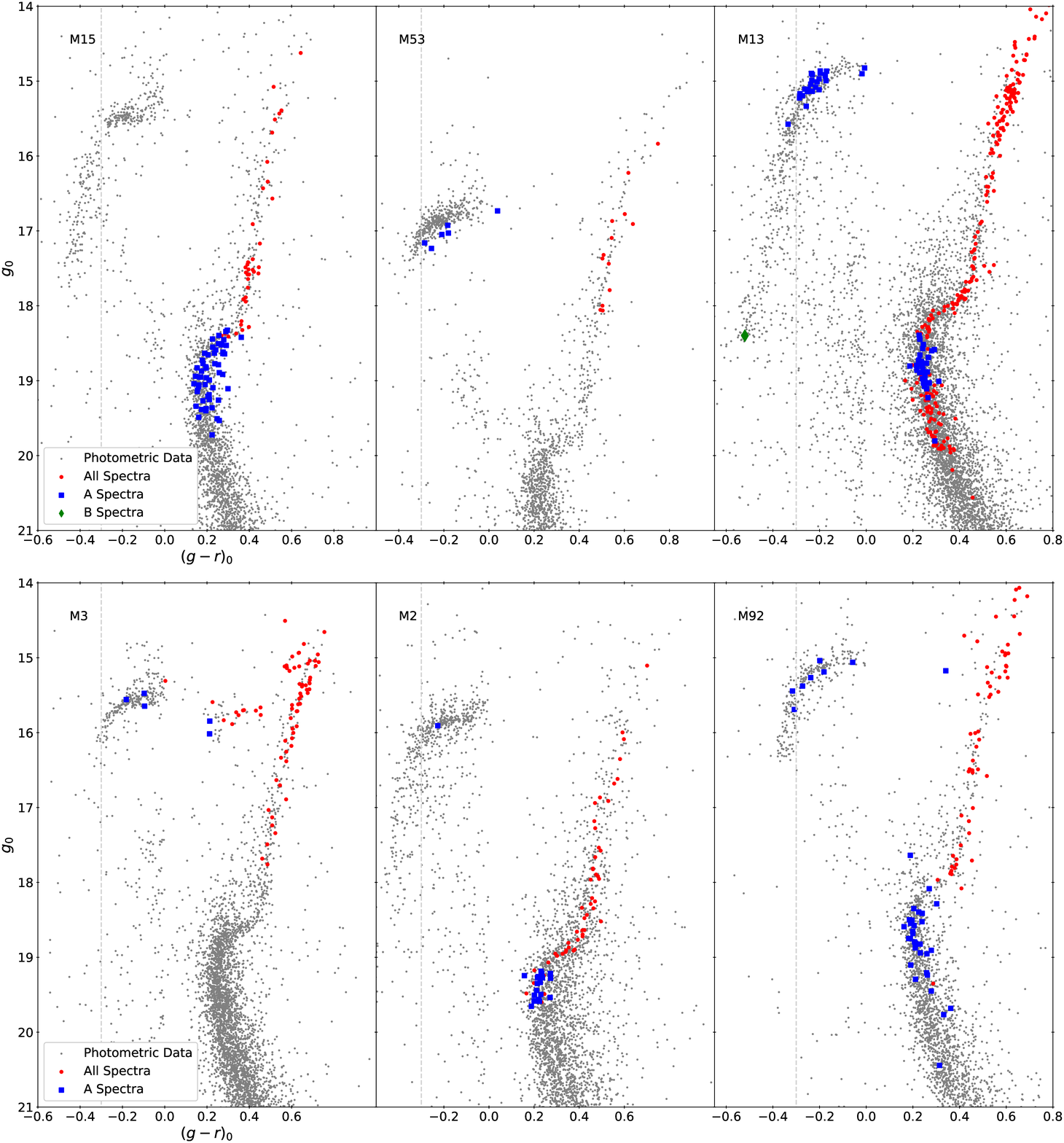}
\caption{
Globular clusters with photometric data as gray dots from \citet{daophot_ref}; stars with color $(g-r)_0 > 0$ have been randomly sampled to show 10\% of their original data. Spectroscopic data collected from \citet{2011_smolinski} and \citet{2016AJ....151....7M} are shown as blue squares (A0 and A0p spectral types), green diamonds (B9 spectral type) and red circles (all other spectral types).
This data shows that all of the HBX stars with spectra have assigned SDSS spectral types of A0, A0p or B9. Many turnoff stars also have A0 or A0p designations, but inspection of the individual spectra shows that these stars have very poor signal-to-noise ratios and appear to be misclassified. 
}
\label{fig:M15_a0ps}
\end{figure}

Figure \ref{fig:M15_a0ps} displays the color-magnitude diagrams for M15, M53, M13, M3, M2 and M92 with the photometric data as small gray points (10\% of the data points for stars with $(g-r)_0 > 0$). The remaining clusters presented in \citet{daophot_ref} either did not have a well-populated BHB shoulder or had no verified spectroscopic data. Stars with early A type spectral classifications are shown as blue squares, one B9 star is shown as a green diamond, and the remaining stars with spectra are shown as red circles.
There are many spectra of horizontal branch stars and a few spectra of HBX stars in these clusters. 
The spectral types of the stars on the HBX and in the blue end of the horizontal branch are all early A (A0/A0p) or very late B (B9).
Also apparent is the large number of early A type stars found in the turnoff region that are redder than our cutoff.
Visual examination of these turnoff star spectra led us to conclude that these much fainter stars were misclassified in the SDSS pipeline; the spectra were noisy and did not resemble the A0 and A0p templates. 

In each cluster, our color cut of $(g-r)_0 < -0.3$ includes only HBX stars, some possible blue stragglers, and white dwarfs. White dwarfs are easily removed from our dataset because of their characteristic spectra. The spectral types for the stars in or near the HBX in these clusters are consistent with the spectral types of candidate HBX stars found in our dataset, reinforcing the idea that we have in fact identified field analogues of HBX stars. Also note that there are very few blue straggler stars that are bluer than our color cutoff.

\begin{figure}
\centering
\includegraphics[width = 0.5\linewidth]{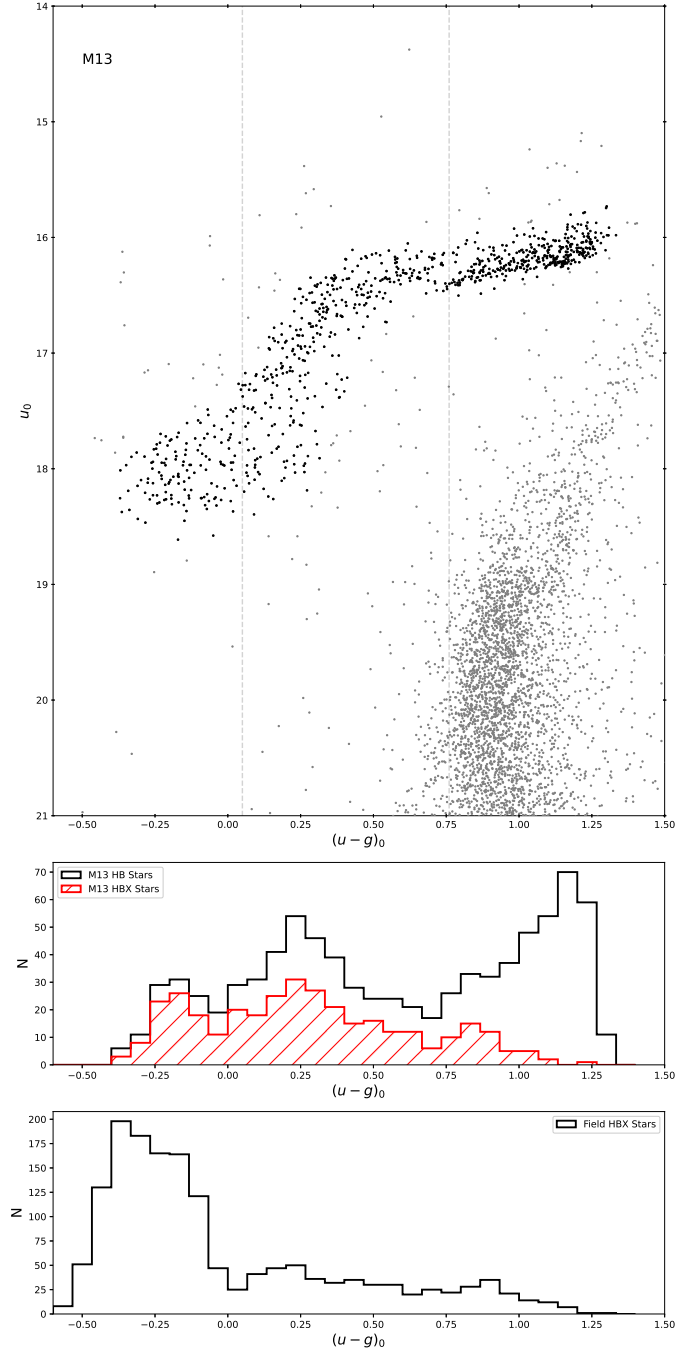}
\caption{
Calibration of stellar types to M13. The M13 CMD (top) shows stars from \citet{daophot_ref}. Stars that were selected by eye to be part of the horizontal branch are shown as black points. Vertical dashed lines indicate the calculated positions of the Grundahl jump at $(u-g)_0 \sim 0.76$, and the Momany jump at $(u-g)_0 \sim 0.05$ in SDSS filters. A histogram of the $(u-g)_0$ colors of the horizontal branch stars is shown in the middle panel. The positions of the Momany and Grundahl jumps are evident as minima at $(u-g)_0\sim 0.0$ and $(u-g)_0=0.75$ - close to the positions at which they were expected. The red histogram shows the portion of the M13 horizontal branch stars that are within our color selection bounds ($(g-r)_0<-0.3$). Note that this color cut eliminates most of the stars redward of the Grundahl jump. The lower panel shows the distribution of $(u-g)_0$ colors in our sample of field HBX stars. The majority of the stars in our sample are bluer than the Momany jump.
}
\label{fig:colorhist}
\end{figure}

To further connect our dataset to stars of known stellar type, we identify the horizontal branch and HBX stars in M13, along with the Grundahl and Momany gaps, and compare the color distribution of actual HBX stars with the color distribution of our field stars. M13 was chosen because its HBX extends the farthest on the blue end, and is the only one of the six clusters known to host EHB stars \citep{1999ApJ...524..242G}. Figure~\ref{fig:colorhist} shows the $u_0$ vs. $(u-g)_0$ CMD. A histogram of the $(u-g)_0$ colors of the stars that were selected by eye to be horizontal branch stars (black points in the top panel of Figure~\ref{fig:colorhist}) is shown by the black line in the middle panel of Figure~\ref{fig:colorhist}. The red line in the middle panel shows the $(u-g)_0$ colors for the subset of the M13 horizontal branch stars with $(g-r)_0<-0.3$, which is the selection criterion for our field HBX stars.

To better interpret this figure, we will estimate the locations of the Grundahl and Momany jumps in SDSS filters. \citet{2002ApJ...576L..65M} put the Grundahl jump at $U-V \sim -0.30$ and the Momany jump at $U-V \sim -1.0$. \citet{1999A&A...343..904C} place the Grundahl jump at $B-V \sim 0.0$. We estimate the Momany jump to be at $B-V \sim -0.15$ from Figure~1 of \citet{2002ApJ...576L..65M}. Therefore, the Momany jump is expected to be at $U-B \sim -0.85$ and the Grundahl jump is expected to be at $U-B=-0.3$. Using the \citet{2005AJ....130..873J} conversion from $U-B$ to $(u-g)$ for stars with $R_c-I_c<1.15$ and $U-B<0$, the Momany jump should be at $(u-g)_0\sim 0.05$ and the Grundahl jump should be at $(u-g)_0 \sim 0.76$. These approximate positions for the jumps are indicated as vertical dashed lines in Figure~\ref{fig:colorhist}.

The apparent horizontal branch jumps in the top panel of Figure~\ref{fig:colorhist} and the corresponding gaps in the black histogram in the middle panel are slightly blueward of the estimates from our transformation. The difference is within the errors of the transformation, but it is unsettled as to whether the difference is caused the by the transformation, an actual shift of the gaps of this cluster, or small irregularities in the \citet{daophot_ref} photometry. The correspondance between the gaps and the jumps indicates that the M13 horizontal branch stars with $(u-g)_0 \lesssim 0$ are EHB stars, while those with $0 \lesssim (u-g)_0 \lesssim 0.7$ are what we are calling BHB tail stars between the Momany and Grundahl jumps, and those redder than $(u-g)_0>0.7$ are the normal BHB stars that are typically used to trace Galactic structure. The position of the jumps is not exact. In the lower panel of Figure~\ref{fig:colorhist}, there is a gap in the $(u-g)_0$ distribution of field HBX stars at exactly the position at which we expected to find the Momany jump, $(u-g)_0=0.05$. In addition, there are very few stars in our sample that are redder than the Grundahl jump. 

For the remainder of this paper, we will assume that all of the sample stars bluer than $(u-g)_0=0.05$ are EHB stars, and sample stars redder than $(u-g)_0=0.05$ will be referred to as BHB tail stars. The two populations together are identified as HBX. Note that our sample primarily consists of EHB stars; 1088 of 1579 stars (69\%) are bluer than $(u-g)_0=0.05$.

\subsection{Distance Calibration}

\begin{figure}
\centering
\includegraphics[width=\linewidth]{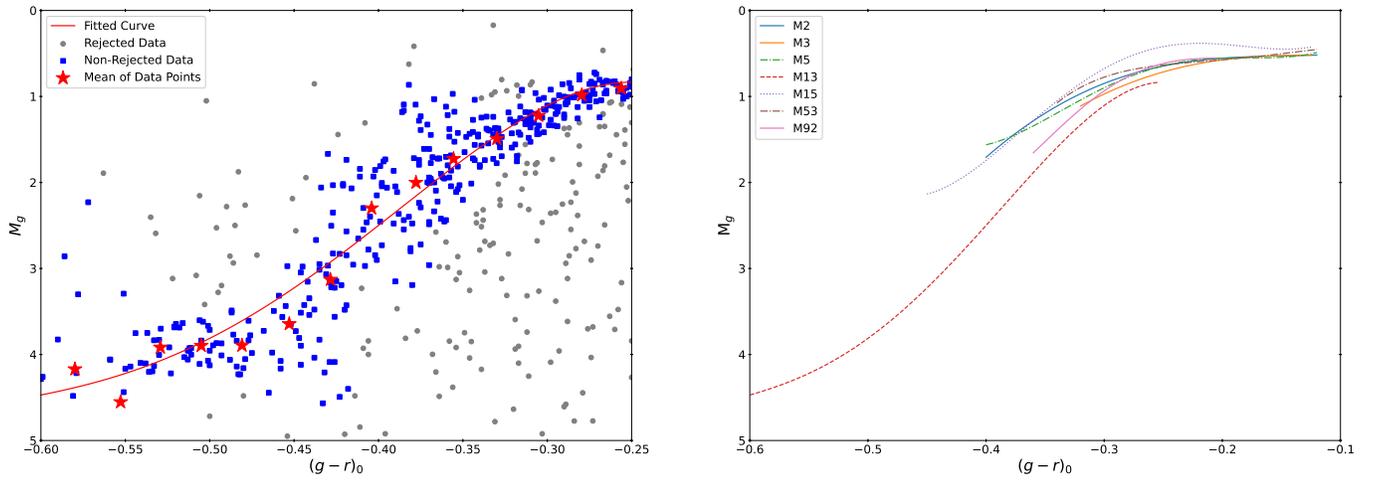} 
\centering
\caption{
\textit{Left:} A polynomial fit to M13's BHB tail.
This was done by taking the average M$_g$ for stars in bins of width 0.05, after iteratively removing data that is further then two standard deviations away.  The stars that were used to compute the mean are shown in blue, and the mean values in each bin are shown as red stars.
The best fit for the data is shown by the red line.
\textit{Right:} Polynomial fits for six other globular clusters' BHB tails, showing the  trend each BHB tail follows is consistent.
}
\label{fig:m13_ehbfit}
\end{figure}

In order to estimate the distances to HBX stars, we derive the absolute magnitudes for known HBX stars as a function of $(g-r)_0$ color. 
Figure \ref{fig:m13_ehbfit} shows a $4^{\rm{th}}$ order polynomial fit to M13's HBX.
M13 was selected because its BHB tail which spans the largest range of $(g-r)_0$ of the six globular clusters studied.
The distance to the globular cluster is 8.4 kpc \citep{harris_catalog}.
When fitting the data, the mean and standard deviation was used to reject data outside of $2 \sigma$ iteratively until the mean was a constant. This analysis was carried out in color bins of size 0.05. 
A polynomial fit, found using an ODR package \citep{2020SciPy-NMeth}, yielded: 
$$M_g =17.36 + 172.43(g-r) +612.74(g-r)^2 + 855.10(g-r)^3 + 421.98(g-r)^4.$$
This fit to the HBX in M13 is shown in the left panel of Figure \ref{fig:m13_ehbfit}.
The average deviation for the actual stars from the line of best fit was 0.45, which gives us an estimate of the error in the derived absolute magnitudes.

To further show that this fit approximates well the absolute magnitude of the HBX, fits were made to six other globular clusters (right panel of Figure \ref{fig:m13_ehbfit}). These clusters have BHB tails that overlap in $(g-r)_0$ with M13, allowing for the absolute magnitudes to be compared. 

This polynomial fit was then used to estimate the distance to M53 and M92, using BHB tail stars with a color in the range $(g-r)_0 < -0.25$, the upper bound to which the HBX was fit.
These are the only clusters (other than M13) with horizontal branch spectra that are near our color cut of $(g-r)_0 < -0.3$. 
In M53 there were two spectra in this cut with calculated distances of 17.2 kpc and 19.0 kpc, for an average distance of 18.1 kpc and a standard deviation of 1.2 kpc. In \citet{harris_catalog}, the distance for M53 is given as 17.9 kpc, which is within the errors of the distance measured from BHB tail stars. The distances measured for the two BHB tail stars differ from the published distance to the cluster by 4\% and 6\%, which is well within the $\sim$20\% error expected if the absolute magnitude is measured to an accuracy of 0.45 magnitudes.
M92 had three stars that fall within this color cut. From the apparent magnitudes of these stars, we derive a distance to M92 of 6.7, 7.9 and 7.8 kpc for an average of 7.5 kpc with a standard deviation of 0.9 kpc. The distance to M92 is 8.3 kpc \citep{harris_catalog}, which is different from each of the three distance measurements by 19\%, 5\%, and 6\%, respectively. For both clusters that had spectra in the color range within which the fit is valid, the distances calculated were within the expected error range of approximately 20\%.

The only globular cluster in our sample in which a spectrum of an EHB star was obtained is M13. The B9 spectral type is the correct temperature for EHB stars, but the spectral classification should have been sdB \citep{1974ApJS...28..157G}. Although it is a bit circular, our distance relation provides an estimate of the distance of 7.6 kpc to that star. This reproduces the distance to M13 within 10\%.

\section{Exploring HBX Substructure in the HAC Region}
\subsection{6D Phase Space}

Using radial velocities and angular positions from SDSS, proper motions from \textit{Gaia} DR2 and distance estimates based on the calculated absolute magnitude of each candidate HBX star, full 6D phase space information was calculated using the transforms outlined in \citet{JohnsonSoderblom_paper}.
We then transform the result to a right-handed, Galactocentric convention where the Sun is at $X=-8$ kpc, and $Y$ is in the direction of the solar motion. $V_X, V_Y,$ and $V_Z$ are Galactocentric velocities that are positive in the direction of positive $X, Y,$ and $Z,$ respectively. This is the convention that will be used throughout this paper.
SDSS DR14 was combined with \textit{Gaia} DR2 by matching stars within one arcsecond of each other in position using TOPCAT \citep{topcat_citation}. 

\begin{figure}
\centering
\includegraphics[width = \linewidth]{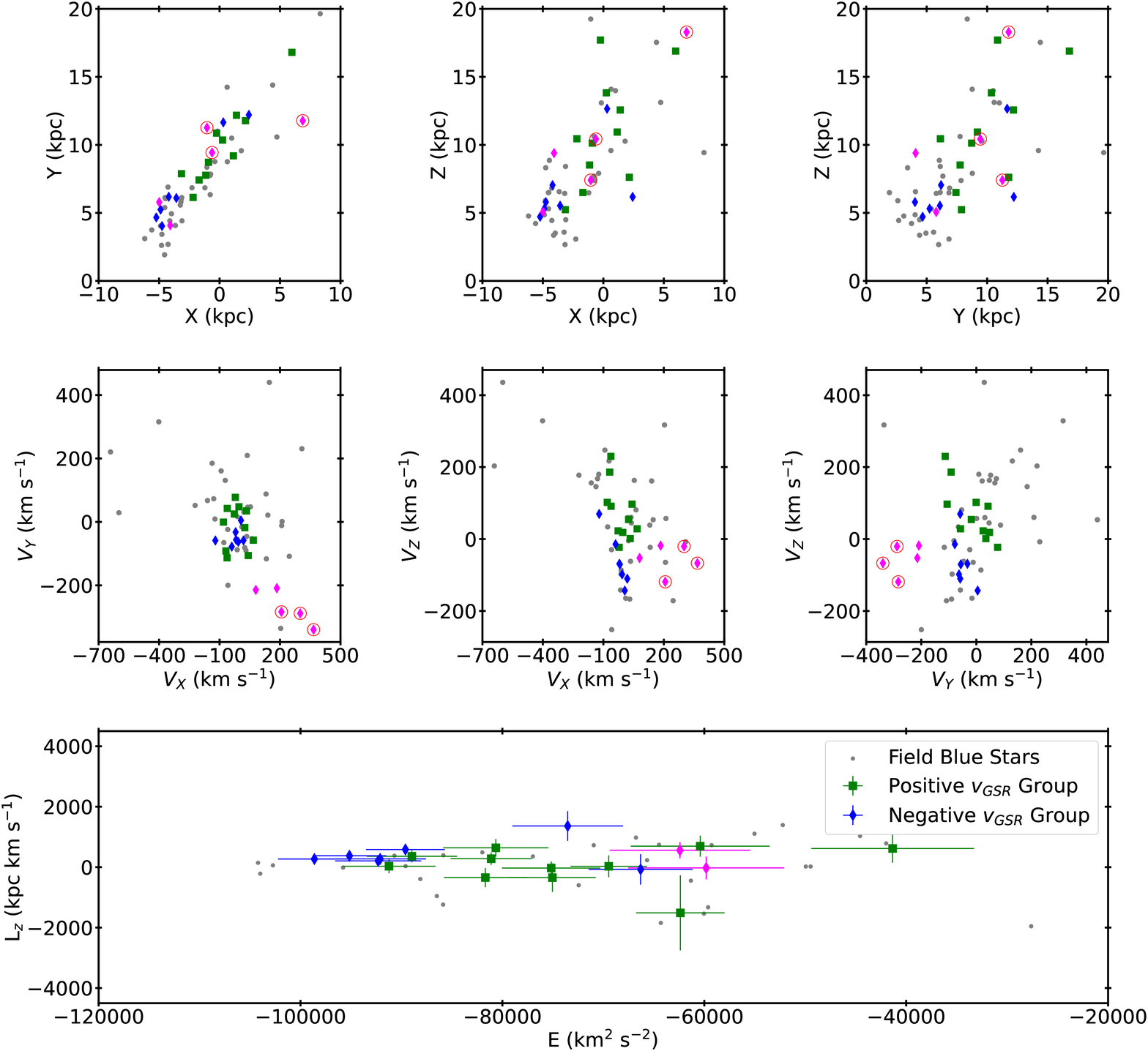}
\caption{
6D phase space information for candidate HBX stars in the overdensity identified in Figure \ref{fig:ra_dec_plots}. 
Stars in the $V_{GSR} = -90$ km s$^{-1}$ group are shown in diamonds. Stars in the $+30$ km s$^{-1}$ are shown in green squares, and stars that are in the boxed overdensity in Figure \ref{fig:ra_dec_plots} that do not belong to either velocity group are shown as filled gray circles. The stars in the more significant negative velocity group appear to separate into two groups in 3D velocity; the stars with large velocities are shown in magenta and those with more typical halo velocities are shown in blue.
XYZ Galactocentric positions are shown in the top panels, and their Galactocenteric velocities, in a right-handed coordinate system, are shown in the second row of panels.
	The bottom panel shows the energy and angular momentum for these stars, as calculated using the same Galactic potential used in \citet{tom_paper}. Three stars shown as circled magenta diamonds have very high ($\sim 20,000$ km s$^{-1}$) energies and very low ($\sim -3500$ kpc km s$^{-1}$) $L_z$, and are therefore not visible on the scale of this plot. In the text we argue that the three high energy stars are in a tidal stream and the majority of the stars in this region (including the blue and green symbols) are part of the VRM.
}
\label{fig:vgsr_plot}
\end{figure}

Figure \ref{fig:vgsr_plot} shows 3D velocities and positions for field HBXs in the identified overdensity, which is located in the sky position of the northern portion of the HAC. The diamonds represent the group of stars at $-90$ km s$^{-1}$, and green squares represent the group at $+30$ km s$^{-1}$. Since the stars in the negative velocity group appear to be separated in velocity, the diamonds are colored to identify the higher velocity stars (magenta) and the stars with velocities in the usual range expected for halo stars (blue).
The bottom panel of Figure \ref{fig:vgsr_plot} shows energy, $E$, and angular momentum in the $Z$ direction, $L_z$, which are integrals of motion for an axisymmetric potential. These parameters were calculated using a potential comprised of a Miyamoto-Nagai disk, a Hernquist bulge, and a logarithmic halo with the parameters from Orphan Stream Model 5 in \citet{potential_2010}. The energies were renormalized by subtracting $60,000$ km$^2$ s$^{-2}$ to match the energy scale in \citet{tom_paper}. 

We start by looking for halo substructure in the negative velocity group, where we see an excess of about eight stars over a reasonable Gaussian background distribution of velocities. Looking first at the magenta diamonds in Figure~\ref{fig:vgsr_plot}, we see that two of them have nearly zero angular momentum. These are the left-most two magenta points in the top panels of Figure~\ref{fig:vgsr_plot}, and the two lower velocity stars in the velocity panels. The other three magenta points have wildly discrepant energy and angular momentum, which is why they do not appear on the $L_z$ vs. $E$ panel. The measured angular momentum for these stars ranges between $-4400$ and $-3000$ kpc km s$^{-1}$ (negative $L_z$ is corotating with the disk in our right-handed coordinate system) with errors of about $1000$ kpc km s$^{-1}$, and the energy ranges from $700$ to $44,000$ km$^2$ s$^{-2}$ with errors of about $20,000$ km$^2$ s$^{-2}$. Note that the zeropoint of potential energy (and therefore the total energy) is set arbitrarily so positive energies do not indicate stars that are unbound from the Milky Way; in the next section we will fit an orbit with an apogalacticon of 28.7 kpc to these stars. These three stars are much further from the Galactic center and have much larger velocities than the other stars in the negative velocity group. These three stars could represent a stream-like halo substructure.

We now focus our attention on the blue points in the negative velocity group. Of the seven stars in this group, there are five with very similar positions, velocities, and energies. These stars have nearly zero angular momentum about the $Z$-axis and are located at $(X, Y, Z) = (-4, 5, 6)$ kpc. A tight moving group like this could also represent a stream-like halo substructure. However, the distribution of angular momentum and energy for the stars in the entire HAC region suggests another possibility.  
From an $N$-body simulation of the VRM, \citet{tom_paper} found that there is a large energy dispersion (-100,000 to -60,000 km$^2$ s$^{-2}$) and a small range of angular momentum (roughly within $\pm 500$ kpc km s$^{-1}$ for the VRM).
Since the VRM is believed to populate the VOD, HAC, and Eridanus-Pheonix Overdensity, it seems possible that a large fraction of the stars in the overdensity might be part of this known radial merger. This interpretation is supported by the large number of stars in the original overdensity and its coincident location with the northern part of the HAC. 
There are only five stars in the negative velocity range that are tightly clustered in phase space, while the overdensity of stars in the right panel of Figure~\ref{fig:ra_dec_plots} contains many more stars.
The HAC could be a component of a large radial merger event which is the primary component of the stellar halo. If this is the case, we would expect the majority of the stars in the HAC to belong to this merger debris, and that there would be very little smooth halo background. 

This might also explain why our velocity distribution in Figure~\ref{fig:vgsr_ra} does not look like it has an underlying smooth Gaussian distribution. Even if we could find a substructure of $\sim 8$ stars that explains the negative velocity peak, it would not explain what looks like a factor of two or more stars in the HAC region than in adjacent low latitude regions. We would need all of the stars in both peaks to be in one or more substructures to explain the observed overdensity. If there is more than one substructure, they would have to be clustered in one sky area.

We will henceforth refer to the high velocity (magenta) group of three stars within the negative $V_{GSR}$ group as the high energy group, which we will show is on a stream-like orbit. We will refer to the lower velocity (blue) group of five stars in the negative $V_{GSR}$ group as the low energy group, which we will associate with the VRM.

\subsection{Stream-like and Shell-like Orbits in the Negative Velocity Group}
To further explore the two negative velocity moving groups that were identified in the previous section, we calculated orbit fits. We again used a potential comprised of a Miyamoto-Nagai disk, a Hernquist bulge, and a logarithmic halo with the parameters from Orphan Stream Model 5 in \citet{potential_2010}. 

\begin{figure}
\centering
\includegraphics[width=\linewidth]{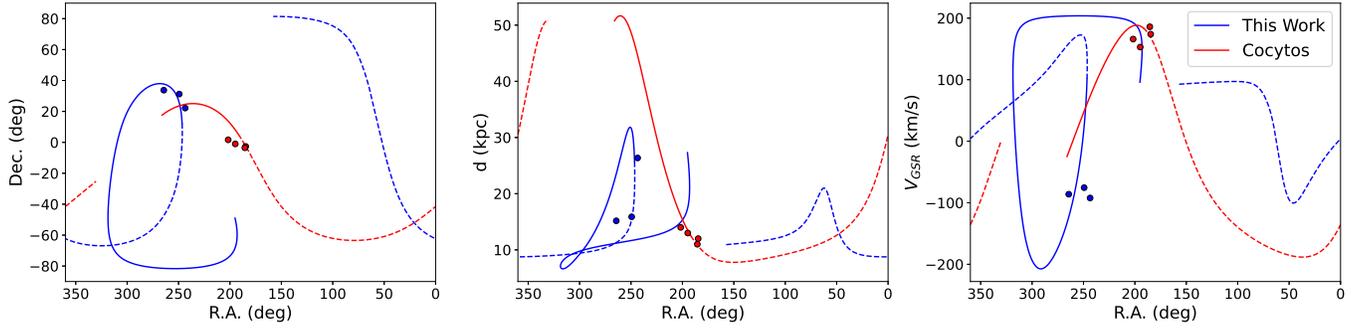}
\caption{
Orbit for the stars in the stream-like negative velocity group. 
Fit orbit of the high energy stars (blue) compared to the fit orbit for Cocytus \citep[red, ][]{tom_paper}. The left panel shows position on the sky for the two orbits, the middle panel shows the orbits' heliocentric distance as a function of R.A., and the right panel shows the line-of-sight velocity of both orbits as a function of R.A. The stars that were used to fit the orbits are plotted in each orbit's respective color.
}
\label{robert_orbit}
\end{figure}

We fit an orbit to the high energy group of three stars using the procedure described in \citet{Willett2009}. This method minimized the sum of the squares of the differences between the model orbit and the observed sky positions, distances from the Sun, and line-of-sight velocities of the observed stream stars. Figure~\ref{robert_orbit} shows the resulting orbit, which has an eccentricity of $e=0.67$, a perigalacticon of 5.6 kpc, and an apogalacticon of 28.7 kpc. The phase space parameters for the fit orbit are $(l,b,d, V_X, V_Y, V_Z) = (27.52^\circ, 37.73^\circ, 25.66$ kpc, $0.01$ km s$^{-1}, 94.51$ km s$^{-1}, 81.88$ km s$^{-1}$).

Also shown in Figure~\ref{robert_orbit} is a comparison between our orbit fit to the high energy stars and the orbit fit by \citet{tom_paper} to the Cocytus stream \citep{Grillmair2009}. 
\citet{tom_paper} identify a group of stars in the vicinity of the VOD that were near the position of the Cocytus stream in that region of the sky and made the tentative connection. 
Their orbit has an eccentricity of $e=0.61$, which is similar to that of our orbit fit, with a perigalacticon of 11 kpc and an apogalacticon of 44 kpc, which are somewhat different from our orbit. 
Both our orbit for the high energy stars and the orbit fit in \citet{tom_paper} are very different from the orbit that \citet{Grillmair2009} fit to his photometric data. \citet{Grillmair2009} identified a faint overdensity of stars that stretches from Virgo [(R.A., Dec.)$=(186^\circ, -3^\circ)$] to Hercules [(R.A., Dec.)$=(259^\circ, 20^\circ)$] at a distance of about 11 kpc.
The stars identified by \citet{tom_paper} are very well aligned with one end of the photometrically identified Cocytus stream, but their velocities make it impossible for the stream to stay at 11 kpc all the way to Hercules. The three high energy stars we identify in this paper are possibly at the right sky position and distance to be associated with the other end of the Cocytus stream, but they are not on the same orbit as the stars in \cite{tom_paper}, and are not along a stream that stays at a distance of $\sim 11$ kpc all the way to Virgo.

\citet{Grillmair2009,Grillmair2014} find six streams (Hermus, Hyllus, Cocytus, Lethe, Styx, and Acheron) that pass through the region of the HAC where we identified an overdensity of HBX stars. Of these, all but Styx, which was found a distance of 45 kpc from the Sun, are within 20 kpc of the Sun. Cocytus and Lethe were found at 11 kpc and 13 kpc, respectively. None of these relatively nearby streams have well-determined orbit fits. Given the complexity of the halo in this region and the possibility that the identified streams might result from the combination of actual streams, we have deferred a study of this and related stream-like substructures to a future paper. However, we do show that our BHB tail stars have the potential to trace halo substructure.

\begin{figure}
\centering
\includegraphics[width=.75\linewidth]{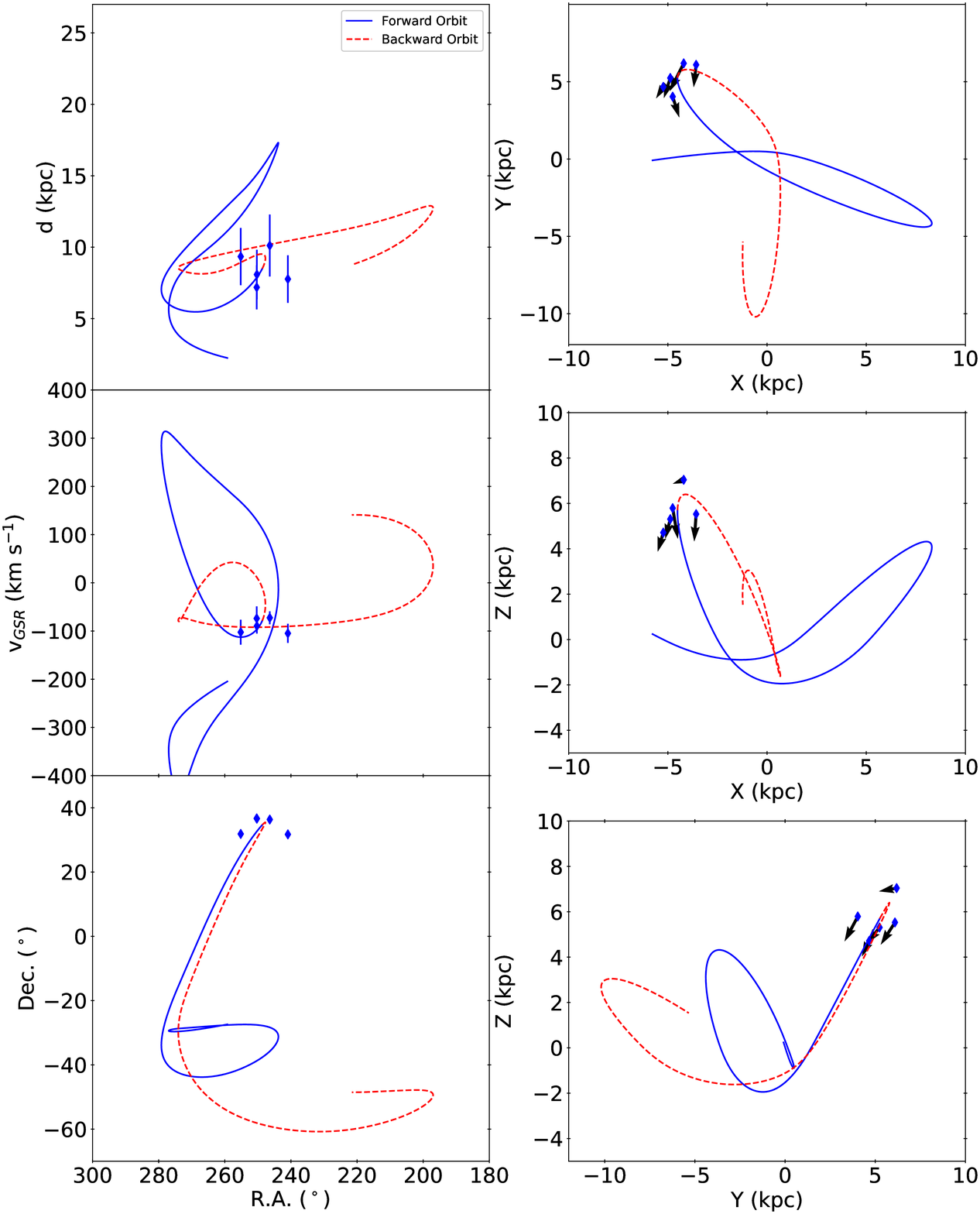}
\caption{
Orbit for the stars in the low energy portion of the negative velocity group. 
The left panels show the distance, line-of-sight velocity in the Galactocentric rest frame, and the declination as a function of right ascension. The right panels show Galactocentric positions of the orbit compared to the positions of the candidate HBX stars, including velocity vectors. Note this debris is on a very radial orbit, which passes very close to the Galactic center.
}
\label{new_orbits}
\end{figure}

Figure~\ref{new_orbits} shows the orbit fit to the low energy group of five stars that have near zero $L_z$. Since the phase space coordinates for these stars are similar, the orbit parameters were determined by the average $X, Y, Z, V_X, V_Y,$ and $V_Z$ coordinates for the stars: $(l, b, d, V_X, V_Y, V_Z) = (56.43^\circ, 42.13^\circ, 8.50$ kpc$, -13.18$ km s$^{-1}, -57.58$ km s$^{-1}, -72.43$ km s$^{-1})$. This orbit is highly radial (eccentricity $e=0.82$), and passes very close to the Galactic center (perigalacticon of 1.0 kpc, and apogalacticon of 10.7 kpc). The orbit is bent by gravity from the disk, so that consecutive apogalacticons are on the north side of the disk. The highly radial orbit is consistent with our initial guess that this moving group might be associated with the VRM. In the next section, we will use $N$-body simulations along this orbit to make more detailed comparisons with the VRM.

\subsection{Connecting HAC BHB tail Stars to the VRM}
An $N$-body simulation of a dwarf galaxy was evolved along the orbit of the low energy HBX group with negative line-of-sight, Galactic standard of rest velocity, $V_{GSR}$. The simulated dwarf galaxy was created using parameters from \citet{tom_paper}, with the exception that the scale radius was reduced to 0.2 kpc and the evolution time was increased to match the merger time estimate from \citet{tom_new_paper}. The progenitor was modeled as a 10,000-body Plummer sphere with a mass of $10^7$ M$_\odot$, a scale radius of 0.2 kpc, and an evolution time of 2.4 Gyr. The small mass was selected by \citet{tom_paper} to trace this group and is not representative of the dwarf galaxy progenitor, which is likely orders of magnitude more massive. 

\begin{figure}
\centering
\includegraphics[width=\linewidth]{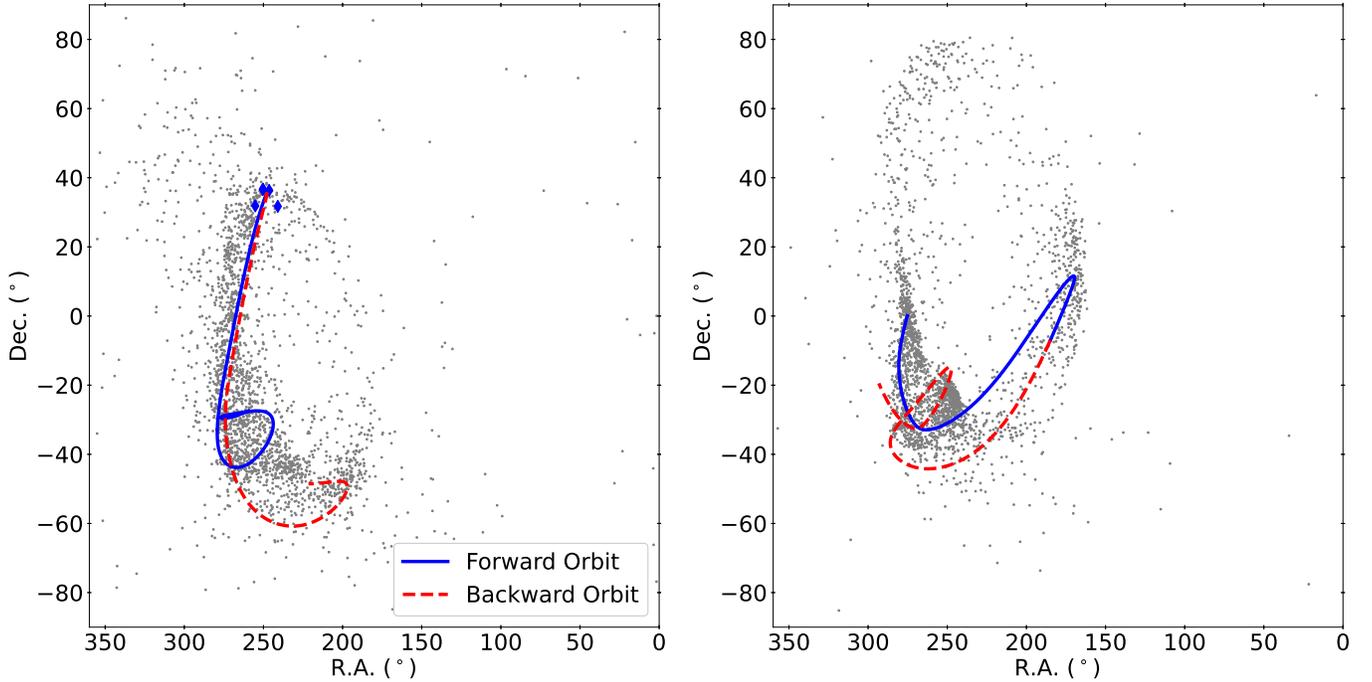}
\caption{
$N$-body simulation for the negative $V_{GSR}$ group of HBX stars \textit{(left)} compared to a simulation of Group D \textit{(right)}, a lower energy component of the VRM evolved along the orbit found in \citet{tom_paper}. Both the simulations used a progenitor mass of 10$^7$ M$_\odot$, a scale radius of 0.2 kpc, 10,000 bodies and an evolution time of 2.4 Gyr. 10\% of the simulation bodies are shown. Comparison of the simulation for our HBX stars to that of the low energy VRM debris, which  shows debris in similar regions of the sky and on similar orbits, suggests that they are related.
}
\label{nbody_fig}
\end{figure}

The left panel of Figure \ref{nbody_fig} shows 10\% of the total bodies simulated. These results can be compared to $N$-Body simulations with identical parameters (right panel of Figure \ref{nbody_fig}) evolved along the \citet{tom_paper} orbit for Group D, which represents a low energy portion of the VRM in the VOD.  Group D was selected for comparison since its stars have energies similar to those of our HBX stars. 
The simulation along the orbit of the HBX stars was run by starting a single point at the position of the HBXs, running it backwards through the potential for 2.4 Gyr, placing a simulated dwarf galaxy at that position, and then running the simulation along the forward orbit for 2.4 Gyr.
The simulation of Group D was run by starting a single point at the position of the Group D stars in the VOD, running it backwards through the potential for 2.4 Gyr, placing an identical simulated dwarf galaxy at that position, and then running the simulation along the forward orbit for 2.4 Gyr.
In R.A. and Dec., the simulations for both Group D and our HBXs place the stars in similar locations. In particular there are many stars in both simulations around $(250^\circ, -30^\circ$). The simulation in the left panel additionally puts stars in the region of the sky where we see an excess of HBXs at around (R.A., Dec.)$= (250^\circ, 30^\circ)$, and the right panel puts stars in the region of the VOD [(R.A., Dec.) $=(190^\circ, 0^\circ)$].

\begin{figure}
\centering
\includegraphics[width=\linewidth]{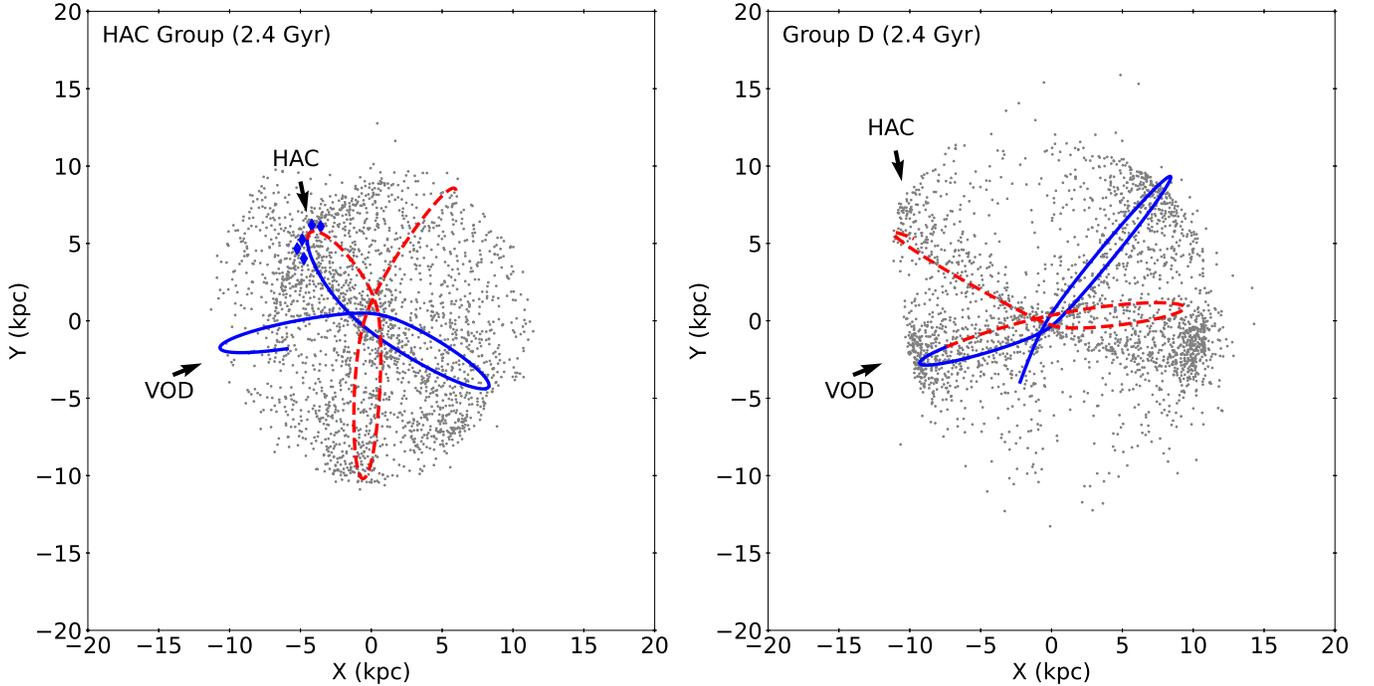}
\caption{
Further comparison of $N$-body simulations from the negative $V_{GSR}$ group (\textit{left})  and Group D from \citet{tom_paper} (\textit{right}). 
The Galactrocentric X and Y positions for both simulation, along with their corresponding orbits integrated for 0.3 Gyrs forwards and backwards. In both panels, the solid blue line represents the forward orbit, and the dashed red line represents the backwards orbit. This shows consistency between the two $N$-body simulations and orbits for stars in the VRM that share the same energy. Moving the orbit in the left panel forward in time leads to the VOD, while moving backwards in time in the right panel would place stars in the HAC region. This suggests that the stars found in \citet{tom_paper} are further along their orbit but share the same orbit and have the same progenitor.
}
\label{nbodyXY_fig}
\end{figure}

The connection between the low energy group and Group D in the VRM is apparent from Figure \ref{nbodyXY_fig}, which shows these same $N$-body simulations in the $X-Y$ plane. In this figure, the orbits of both simulations are plotted over the simulation data. The forward orbits for both are shown as blue solid lines, and the backwards orbits are shown as red dashed lines. Going forward in time on the left panel by 0.3 Gyr, which is approximately the galaxy crossing time for stars on this orbit, places stars directly where the orbit in the right panel begins, at the location of the VOD. Similarly, going backwards in time from the right panel by 0.3 Gyr places stars at the location where we found the negative $V_{GSR}$ group.
The time difference between the two simulations results from the fact that each simulation was started from a different position on the orbit but evolved the same time.
This suggests that both the negative $V_{GSR}$ group and Group D from \citet{tom_paper} are on the same orbit; Group D is just further along the orbit. 

\section{HBX stars in the halo and the disk}
We have argued that one piece of the HBX substructure in the HAC, and likely much of the HBX overdensity in the HAC, is part of the VRM. But the VRM is expected to be a significant halo substructure that populates all parts of the halo. In this section we will analyze the HBX sample as a whole. We will identify stars that are likely disk HBX stars and likely halo HBX stars, and discuss the distribution of eccentricities in the stellar halo.

\begin{figure}
\centering
\includegraphics[width = \linewidth]{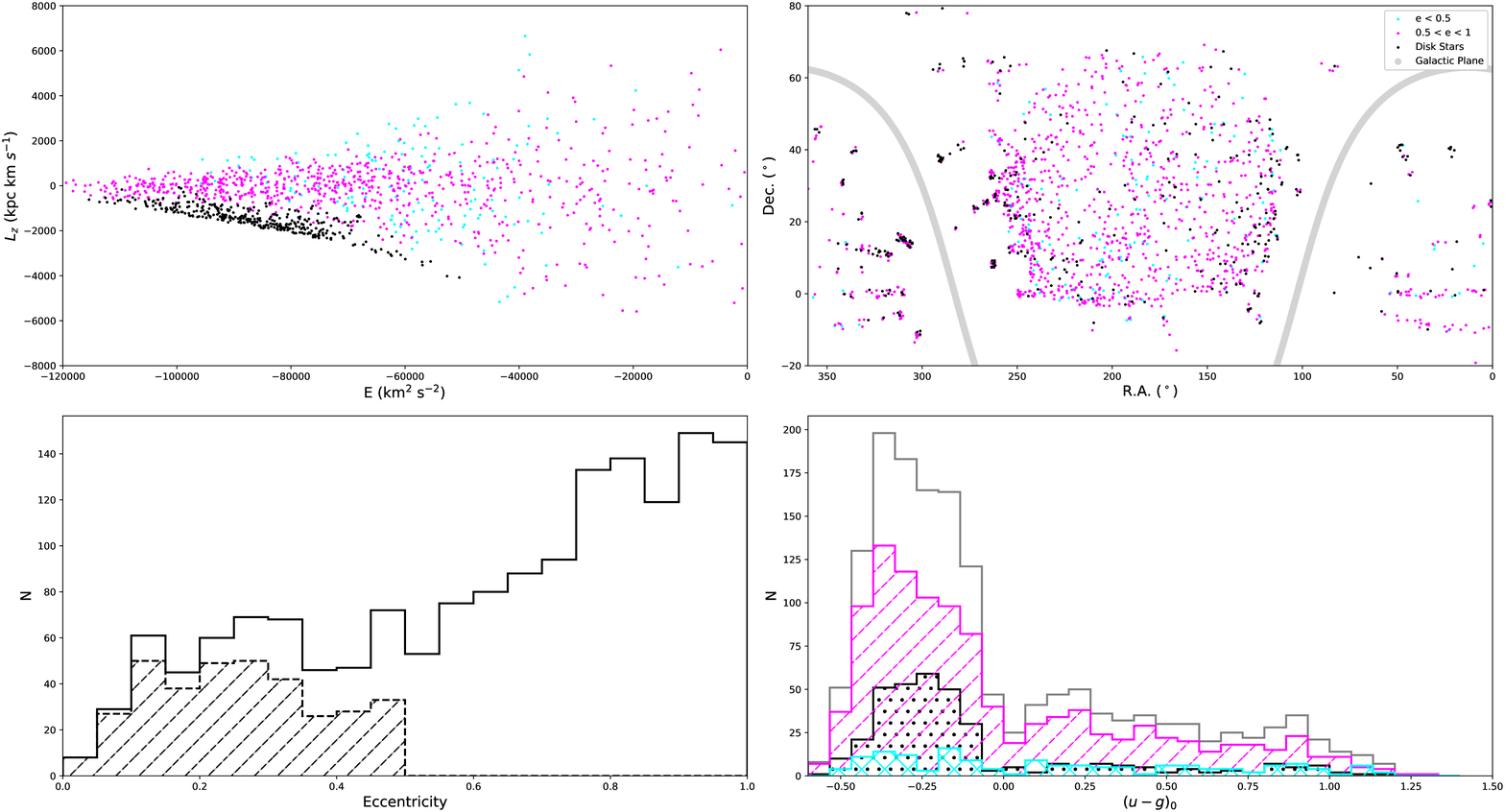}
\caption{Separation of candidate HBX stars into disk stars and halo stars. \textit{Top left:} We show angular momentum around the $Z$ axis as a function of energy for all 1579 presumed HBX stars. Those with eccentricities greater than 0.5 are shown in magenta (1074 stars). Stars with eccentricities less than 0.5 and a combination of energy and angular momentum that suggests they are disk stars are shown in black (352 stars). The rest of the stars with eccentricity less than 0.5 are shown in cyan (153 stars). These eccentricities were calculated by creating an orbit for each star, integrating forwards and backwards 0.5 Gyr, and using Galpy's eccentricity function \citep{galpy_cite} to calculate the eccentricity of each star. \textit{Bottom left:} A histogram of eccentricities of HBX stars (black line) shows that they are predominantly on radial (high eccentricity) orbits. This is more apparent if the presumed disk stars (gray dashed line) are removed from the sample. 
\textit{Top right:} We show the R.A. vs. Dec. for our sample stars, using the same colors as in the top left panel. The location of the Galactic plane is indicated with a gray line. Note that the presumed disk stars (black) are preferentially located at low Galactic latitudes, as expected. Also note that there are very many high eccentricity halo stars that hug the low Galactic latitudes between the HAC $(250^\circ, 30^\circ)$ and the VOD $(190^\circ, 0^\circ)$. These stars are likely HBX stars in the VRM. \textit{Bottom right:} We show the $(u-g)_0$ color distribution of HBX stars in our sample, separated by component, using the same colors as used in the other panels. Note that the disk stars (black) have the highest fraction of EHB stars, which are bluer than the Momany jump at $(u-g)_0=0.05$. The high eccentricity halo stars have a slightly lower fraction of EHB stars. And the low eccentricity halo appears to be drawn from a different stellar population that has a much lower fraction of EHB stars.
}
\label{allEHB}
\end{figure}

Figure \ref{allEHB} shows the 1597 candidate HBX stars, spread over the entire SDSS footprint. We calculated an orbit for each star individually using the Galpy python package \citep{galpy_cite}. These orbits were integrated both forwards and backwards in time for 0.5 Gyr. Then, we used the Galpy built-in numerical orbital eccentricity method to estimate the orbital eccentricity of each star's orbit. The lower left panel (solid black line) shows the eccentricities for all of the stars in the sample; 68\% of the stars have an eccentricity above 0.5. Our expectation is that disk stars will have an eccentricity less than 0.5 and VRM stars will have an eccentricity greater than 0.5, so we first separate the stars into these two samples. In the top left panel of Figure~\ref{allEHB}, we show the angular momentum in the $Z$ direction as a function of energy. High eccentricity stars are shown in magenta. The low eccentricity stars are separated into two groups: stars that hug the low $L_z$ edge of the distribution for their energy (negative $L_z$ in our right-handed coordinate system is corotating with disk stars) are assumed to be disk stars (black points) and the rest are assumed to be low eccentricity halo stars (cyan points). There are 352 stars in the disk sample. If one looks at the region of positive $L_z$ that is located at the symmetrical position on the other side of $L_z=0$ from the disk stars, there are very few (of order 20) stars with low eccentricity in the halo. We show the distribution of eccentricities of disk stars in the lower left panel of Figure~\ref{allEHB}. Note that the disk stars dominate the low eccentricity stars; a very small fraction of the stars in the sample are low eccentricity halo stars.

In the right panel of Figure~\ref{allEHB} we show the sky positions of the stars in the disk and halo samples. The disk stars (black points) are found predominantly at low Galactic latitudes. Note, however, that there are also a large number of high eccentricity halo stars at low Galactic latitudes between the HAC region, (R.A., Dec.)$= (250^\circ, 30^\circ)$, and the VOD, (R.A., Dec.)$= (190^\circ, 0^\circ)$. In the top left panel one sees that the low eccentricity halo stars are spread across the allowed region of the $L_z$ vs. $E$ diagram, but there is a large concentration of high eccentricity halo stars near $L_z=0$ and with energies less than $E=-60,000$ km$^2$ s$^{-2}$. These high eccentricity stars populating the region from the HAC to VOD are likely members of the VRM, and are a large fraction of the HBX sample. From Figure~\ref{fig:ra_dec_plots} it is evident that the low latitude halo structure extending south from the HAC is comprised of brighter stars, and is therefore closer to the Sun. These stars were not originally identified as an overdensity because an increase of stars towards the plane, particularly towards the Galactic center, is generally expected to be contamination from disk stars. Note that many VRM halo stars on high eccentricity orbits will populate other regions of the sky. For example, the $N$-body simulation in the right panel of Figure~\ref{nbody_fig} puts stars near (R.A., Dec.)$= (190^\circ, 10^\circ)$.

Figure~\ref{allEHB} shows that we can identify likely HBX stars in the disk and the halo; the halo population appears to be predominantly related to the VRM, but a portion of the stars are unlikely to be associated with the VRM due to their eccentricity, energy, or location in the sky. The samples identified here can be used to trace old stellar populations in the Milky Way. In future work we plan to combine these stars with other sparse tracers of these populations, such as BHB and RR Lyrae stars, to map the structure of the old stellar population in our galaxy.

With the HBX sample now divided into Milky Way components, we can explore the morphology of the horizontal branch as a function of component. The lower panel of Figure~\ref{fig:colorhist} shows the $(u-g)_0$ distribution for the disk, high eccentricity halo, and low eccentricity halo. We notice that there are very few EHB stars (bluer than $(u-g)_0=0.05$) in the low eccentricity halo. Seventy-four of 154 low eccentricity halo stars (48\%) have EHB colors. This distribution is reasonably comparable to the distribution in M13, which is thought to have a comparatively blue horizontal branch compared to other clusters. On the other hand the disk has a very high fraction of EHBs; 283 of the 351 presumed disk stars (81\%) are EHBs. Clearly this indicates a different population. The stars in the high eccentricity halo, that could be associated with the VRM, also have a high fraction of presumed EHB stars, though not as high a fraction as the disk. 731 of the 1074 stars in the high eccentricity halo (68\%) have EHB colors.

The EHB fraction in each component might tell us something interesting about stellar populations or help understand the orgins of EHB stars themselves. EHB stars (and field sdB stars) have measured masses of about $0.5 M_\odot$ \citet{1986A&A...155...33H}; to reach the helium core burning stage with this mass, they are thought to have experienced significant mass loss, possibly through the transfer of mass during binary evolution \citep{2002MNRAS.336..449H, 2003MNRAS.341..669H}. The large fraction of EHB stars in disk might tell us the disk has a high fraction of close binaries.

Whatever the formation mechanism is for HBX and EHB stars, the fact that they follow a reasonably well-defined color-absolute magnitude relation and are associated with specific populations allows us to use them to trace halo substructure. Since the EHB stars strongly favor some stellar populations over others, they could help to separate different halo populations. If the VRM is the only component of the halo that has a large number of EHBs, then the distribution of EHB stars in the halo would trace the VRM density, for example.

\section{Conclusion}
In this paper we outline a method for identifying HBX stars in the field, derive a formula to estimate their absolute magnitudes from color, and use identified HBX stars to explore the 6D phase space distribution of stars in the Milky Way. We find many likely members of the VRM, which we believe is predominantly the same merger remnant as the \textit{Gaia} Sausage and \textit{Gaia}-Enceladus structure. We create a set of candidate HBX stars that trace the disk, the halo, and the VRM.

We selected 1579 likely HBX stars from SDSS DR14 by choosing point sources with $(g-r)_0 < -0.3$, and then eliminated white dwarfs, blue main sequence stars and subdwarf O stars identified by spectral template matching. This requires the availability of both photometric and spectroscopic data. We connect the HBX stars to the horizontal branch of M13, and show that the stars blueward of $(u-g)_0 \lesssim 0.05$ are likely EHB stars.

A formula to calculate approximate HBX absolute magnitudes was derived by fitting the absolute magnitude of the HBX stars in M13 as a function of $(g-r)_0$ color.  Other clusters approximately follow this same polynomial fit. We recover distances to globular clusters to an accuracy of about 20\% by this method.

We identify an overdensity of presumed HBX stars with $18 < g_0 < 19$ in the northern HAC region. The velocity distribution in this overdensity is not consistent with a smooth halo distribution, and therefore indicates substructure. We identify three high energy stars that appear to be members of a stream-like halo substructure. We traced orbits for a low energy comoving group of five candidate HBX stars in the direction of the HAC, and show that these stars are less than 0.3 Gyr behind, but are on similar orbits to, the stars in the low energy component of the VRM. An $N$-body simulation evolved along the orbit of these candidate HBX stars showed similar placements of debris across the Milky Way as a simulation of VRM stars of similar energy. 

A large fraction of our sample (1074 stars) are high eccentricity halo stars. A large fraction of these stars are low $L_z$ stars concentrated towards the Galactic plane between the HAC and VOD, and are likely members of the VRM. A small fraction of the stars (153 stars) are low eccentricity halo stars. We also find a significant number of HBX stars (352 stars) with disk-like kinematics, that are preferentially located near the Galactic plane. The morphology of the horizontal branch in these three samples is very different. The disk HBXs have the highest fraction of EHB stars (81\%), the high eccentricity halo has the next highest fraction (68\%), and the low eccentricity halo has the lowest EHB fraction (32\%). These field HBX stars can be used to trace Milky Way substructure.

\acknowledgments
We thank the anonymous referee for encouraging us to explore the morphology of the horizontal branch. We thank V. Belokurov for providing us with the HAC data used in the \citet{belokurov2007} paper. This work was supported by NSF grant AST 19-08653; contributions made by the Marvin Clan, Babette Josephs, and Manit Limlamai; and the 2015 Crowd Funding Campaign to Support Milky Way Research. T.D. was partially supported by a NASA/NY Space Grant fellowship.

Funding for the Sloan Digital Sky Survey IV has been provided by the Alfred P. Sloan Foundation, the U.S. Department of Energy Office of Science, and the Participating Institutions. SDSS-IV acknowledges support and resources from the Center for High-Performance Computing at the University of Utah. The SDSS web site is www.sdss.org.

SDSS-IV is managed by the Astrophysical Research Consortium for the Participating Institutions of the SDSS Collaboration including the 
Brazilian Participation Group, the Carnegie Institution for Science, Carnegie Mellon University, the Chilean Participation Group, the French Participation Group, Harvard-Smithsonian Center for Astrophysics, Instituto de Astrof\'isica de Canarias, The Johns Hopkins University, Kavli Institute for the Physics and Mathematics of the Universe (IPMU) / University of Tokyo, the Korean Participation Group, Lawrence Berkeley National Laboratory, 
Leibniz Institut f\"ur Astrophysik Potsdam (AIP), Max-Planck-Institut f\"ur Astronomie (MPIA Heidelberg), Max-Planck-Institut f\"ur Astrophysik (MPA Garching), Max-Planck-Institut f\"ur Extraterrestrische Physik (MPE), National Astronomical Observatories of China, New Mexico State University, 
New York University, University of Notre Dame, Observat\'ario Nacional / MCTI, The Ohio State University, 
Pennsylvania State University, Shanghai Astronomical Observatory, United Kingdom Participation Group, Universidad Nacional Aut\'onoma de M\'exico, University of Arizona, University of Colorado Boulder, University of Oxford, University of Portsmouth, University of Utah, University of Virginia, University of Washington, University of Wisconsin, Vanderbilt University, and Yale University.

This work has made use of data from the European Space Agency (ESA) mission
{\it Gaia} \linebreak (\url{https://www.cosmos.esa.int/gaia}), processed by the {\it Gaia}
Data Processing and Analysis Consortium (DPAC,
\url{https://www.cosmos.esa.int/web/gaia/dpac/consortium}). Funding for the DPAC
has been provided by national institutions, in particular the institutions
participating in the {\it Gaia} Multilateral Agreement.

\pagebreak

\bibliographystyle{aasjournal}
\bibliography{biblio} 

\end{document}